\documentstyle[12pt,epsfig,multirow]{article}

\def\la{\mathrel{\mathchoice {\vcenter{\offinterlineskip\halign{\hfil
$\displaystyle##$\hfil\cr<\cr\sim\cr}}}
{\vcenter{\offinterlineskip\halign{\hfil$\textstyle##$\hfil\cr<\cr\sim\cr}}}
{\vcenter{\offinterlineskip\halign{\hfil$\scriptstyle##$\hfil\cr<\cr\sim\cr}}}
{\vcenter{\offinterlineskip\halign{\hfil$\scriptscriptstyle##$\hfil\cr<\cr\sim
\cr}}}}}

\newcommand{\be}{\begin{equation}}
\newcommand{\ee}{\end{equation}}
\newcommand{\bea}{\begin{eqnarray}}
\newcommand{\eea}{\end{eqnarray}}

\begin{document}

\begin{flushright}
\texttt{SINP/INO-2010/02}\\
\texttt{RKMVU/PHY-2010/02}\\
\end{flushright}
\bigskip

\begin{center}
{\Large \bf Constraining Scalar Singlet Dark Matter\\
 with CDMS, XENON and DAMA and \\
Prediction for Direct Detection Rates}
\vspace{.5in}

{\bf Abhijit Bandyopadhyay$^\dagger$,
Sovan Chakraborty$^\ddagger$,
Ambar Ghosal$^\ddagger$ and\\
Debasish Majumdar$^\ddagger$}\\
\vskip .5cm
{\normalsize \it $\dagger$ Ramakrishna Mission Vivekananda University,} \\
{\normalsize \it Belur Math, Howrah 711202, India}\\
\vskip 0.2cm
{\normalsize \it $\ddagger$ Saha Institute of Nuclear Physics,} \\
{\normalsize \it 1/AF Bidhannagar, Kolkata 700064, India}

\vskip 2cm

{\bf ABSTRACT}
\end{center}
We consider a simplest extension of the Standard Model (SM) through the 
incorporation of a real scalar singlet and an additional discrete $Z_2$ 
symmetry. The model admits the neutral scalar singlet to be stable and thus,
a viable component of dark matter. We explore the parameter space of the model 
keeping in view the constraints arise from different 
dark matter direct detection experiments 
through WIMP-nucleon scattering. First of all, we have utilised 
the data obtained from CDMS, XENON-10 and XENON-100 collaborations. 
We further constraint the parameter space from the 
DAMA collaboration results (both with and without channelling) and
CoGeNT collaboration results. 
Throughout our analysis, the constraint arises 
due to the observed relic density of dark matter 
reported by WMAP experiment, is also incorporated. 
Utilising all those constraints, on the model parameter space, 
we calculate the event rates and the annual 
variation of event rates in the context of a Liquid Argon Detector experiment.
\newpage
\section{Introduction}
Several cosmological observations like rotation curves of spiral galaxies,
the gravitational micro-lensing, 
observations on Virgo and 
Coma clusters\cite{virgo,coma}, bullet clusters \cite{bullet},
etc. provide indications of  existence of
huge amount of  non-luminous matter or dark matter(DM) in the universe.
The Wilkinson Microwave Anisotropy Probe 
(WMAP) experiment \cite{wmap} 
suggests that about 85\% of the total matter content
of the universe is dark. This constitutes
23\% of the total content of the universe. The rest 73\%
is the dark energy, whereas the remaining 4\% is the known
luminous matter. Nature and identity of the constituents of 
this non-luminous matter is mostly 
unknown. However, the indirect evidences suggest that 
most of them are stable , nonrelativistic (Cold Dark Matter or CDM) 
and Weakly Interacting Massive Particles (WIMPs) 
\cite{Jungman:1995df, Griest:2000kj, Bertone:2004pz, Murayama:2007ek}.
Despite the wide success spectrum of Standard Model (SM), 
explanation of CDM poses a challenge  
to SM as no viable candidate for CDMs 
has been obtained within the framework of SM.
Hence, there are attempts to explain the DM WIMP candidates with 
theories beyond standard model. 

Phenomenology of different extensions of scalar sector of the 
SM had been explored by 
many groups
\cite{McDonald:1993ex,Bento:2000ah,Burgess:2000yq,
Davoudiasl:2004be,Schabinger:2005ei,O'Connell:2006wi,
Kusenko:2006rh,Bahat-Treidel:2006kx,Andreas:2005,Yaguna:2008HD,He:2009YD,Andreas:2008XY,
Cirelli:2009uv}. 
However, addition of  one real scalar singlet to the SM provides
the simplest possible minimal renormalizable extension to the scalar
sector of SM. In addition, invoking a $Z_2$ discrete symmetry under which 
the additional singlet is odd, gives rise to the singlet as a viable 
DM candidate.
In this work, we explore  
the parameter space  of the model to accommodate the results
of different  dark matter direct detection experiments. 
We further predict the event rates and the annual variation of the event rates 
to be observed by the Liquid Argon Detector experiment.
In the experiments for direct detection of DM,
the WIMP scatters off the target nucleus 
of the material of  the detector giving rise to recoil of the nucleus.
The energy of  this nuclear recoil is very low ($\sim$ keV). 
The signal generated by the nuclear 
recoil is measured for direct detection of dark matter.

There are several ongoing experiments for direct dark matter searches. 
Some of them are cryogenic detectors where the detector material 
such as Germanium
are kept at a very low temperature 
background and the nuclear recoil energy is measured 
using scintillation, phonon or ionization 
techniques. The experiments like CDMS 
(Cryogenic Dark Matter Search 
uses Germanium as detector material) at Soudan Mine, Minnesota \cite{CDMS}
use both ionization and phonon techniques.
In phonon technique, the energy 
of the recoil nucleus sets up a vibration of the detector 
material (Ge crystal for CDMS). 
These vibrations or phonons propagate at the surface of the detector crystal 
and excites quasi-particle states at materials used in the pulse pick up device.
Finally the heat produced by these quasi-particle states is converted to 
pulses by SQUID (Superconducting Quantum Interference Device) amplifiers.   
CDMS carries out two experiments - one with Germanium and the other with
Silicon in order to separate the neutron background. As Germanium nucleus
is heavier ($A=73$) than Silicon ($A=28$),
WIMPs  interact with $^{73}$Ge
with higher probability than with $^{28}$Si, but neutron being strongly
interacting will not make any
such discrimination. Thus any excess signal at the $^{73}$Ge
detector over the $^{28}$Si in CDMS will be a possible signature for dark matter.

The DAMA experiment at Gran Sasso 
(uses diatomic NaI as the detector material) \cite{DAMA}, 
uses the scintillation technique for detection of  the recoil energy. 
There are other class of liquid or gas (generally noble gases) detectors 
that measure the recoil energy by the ionization of the detector 
gas. The ionization yield is amplified by an avalanche process and
the drifting of these charge reaches the top (along z-axis) 
where they are collected by the electrodes for 
generating a signal. These types of detectors known as TPC 
(Time Projection Chamber) are gaining lot of interest in present time 
for their better effectiveness and resolution in detecting 
such direct signals of recoil energy from a DM-nucleon scattering. 
As mentioned, they generally use noble gases such as Xenon,
Argon or Fluorine etc. The XENON-10 experiment \cite{XENON10} at Gran Sasso
is a liquid Xenon TPC
with target mass of 13.7 Kg, whereas its upgraded version, the XENON-100 
experiment \cite{XENON100} uses 100 Kg of the target mass.  
The CoGeNT (Coherent Germanium Neutrino Technology)
experiment \cite{cogent} also uses Ge as detector material and designed 
to detect dark matter particles with masses less than
that to be probed in CDMS.
Recently the CoGeNT collaboration has reported an excess of events above the 
expected background \cite{cogent}.
The experiment ArDM (Argon Dark Matter Experiment) \cite{ArDM} plans to use 
1 ton of liquid $^{39}$Ar 
gas for the TPC. There are other experiments that use other techniques 
like PICASSO \cite{picasso} etc. at SNOlab in Canada but here we 
consider  
CDMS, DAMA, CoGeNT and Xenon experiments for the present study.
We restrict the relevant couplings of the scalar dark matter 
by using the bounds on
dark matter-nucleon scattering cross sections from these three experiments 
and further, utilising the WMAP experimental data. 
We predict possible direct detection 
event rate as well as annual variation of event rate in 
liquid ArDM experiment for different possible dark matter masses.

The paper is organized as follows. In Section 2 we briefly discuss 
the model. The CDMS, XENON, CoGeNT and DAMA bounds used to 
constrain the parameter space of the 
model is described in Section 3. In Section 4 
we present the formalism of Direct detection rate calculations and 
compute such rates for liquid Ar detector. Section 5
contains summary and conclusions. 

\section{Singlet extended Standard Model : A brief outline}
The framework of the simplest scalar sector 
extension of the SM involving addition of a real scalar singlet
field to the SM Lagrangian has been 
discussed in detail in \cite{Barger:2007im,OConnell:2006wi}.
In this section we present a  brief outline of the model  
and emphasize on those of its aspects that would be relevant
for discussions to follow. 

The most general form of the potential
appearing in the Lagrangian density for scalar sector of this model
is given by \footnote{We used same notations as used in \cite{Barger:2007im,OConnell:2006wi}.}
\begin{eqnarray}
V(H,S) 
&=& 
\frac{m^2}{2} H^\dagger H 
+ \frac{\lambda}{4} {(H^\dagger H)}^2
+ \frac{\delta_1}{2} H^\dagger H S
+ \frac{\delta_2}{2} H^\dagger H S^2 \nonumber\\
&&
+ \left(\frac{\delta_1 m^2}{2\lambda}\right)S
+ \frac{\kappa_2}{2}S^2 
+ \frac{\kappa_3}{3}S^3 
+ \frac{\kappa_4}{4}S^4
\label{eqpot}
\end{eqnarray}
where $H$ is the complex Higgs field (an $SU(2)$ doublet) and
$S$ is a real scalar gauge singlet that defines our minimal extension
to the scalar sector of SM. The singlet $S$ needs to be stable
in order to be considered as a viable dark matter candidate. 
Stability of $S$ is achieved within the theoretical 
framework of the model by assuming the potential 
to exhibit a $Z_2$ symmetry $S\rightarrow -S$. 
This ensures absence of vertices 
involving odd number of singlet fields $S$ 
($\delta_1 = \kappa_3$ =0).
Using unitary gauge, we define 
\begin{eqnarray}
H &=& \pmatrix {0 \cr \frac{v+h}{\sqrt{2}}}
\end{eqnarray}
where $h$ is the physical Higgs field and 
$v = 246$ GeV is the VEV of the $H$ scalar determined by the parameters
$m$ and $\lambda$ as $v = \sqrt{\frac{-2m^2}{\lambda}}$. 
The mass terms of the two scalar fields $h$ and $S$ are  
identified as 
\begin{eqnarray}
V_{\rm mass} 
&=& 
\frac{1}{2} (M_h^2 h^2 + M_S^2 S^2)
\end{eqnarray}
where, 
\begin{eqnarray}
M_h^2 &=& -m^2 = \lambda v^2/2 \nonumber\\
M_S^2 &=& \kappa_2 + \delta_2 v^2/2
\label{eqms} 
\end{eqnarray}
%
\begin{figure}[t]
\begin{center}
\includegraphics[width=4.5cm, height=4cm, angle=0]{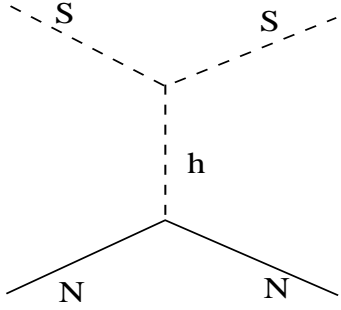}
\caption{\label{fig:fd}
Diagram for singlet-nucleon elastic scattering via higgs mediation}
\end{center}
\end{figure}
The scalar field $S$ is stable as long as the $Z_2$ 
symmetry is unbroken and appears to be a 
candidate for cold dark matter  in the universe. 

In the present work, we investigate the prospect of 
such a candidate in direct detection experiments through
its scattering off nucleon ($N$) relevant for the detector.  The lowest order
diagram for the process has been shown in Fig.\ \ref{fig:fd}.
The cross section corresponding to the elastic scattering ($SN \to SN$) in 
the non-relativistic limit is given by \cite{Burgess:2000yq}
\begin{eqnarray}
\sigma^{\rm scalar}_{N} 
&=&
\frac{\delta_2^2 v^2 |{\cal A}_N|^2}{4\pi} 
\left( \frac{m^2_r}{{M_S}^2{M_h}^4}\right)
\label{eqcross1}
\end{eqnarray}
where, $m_r (N,S)= M_N M_S/(M_N + M_S)$ is the reduced mass 
for the target nucleus of the two body scattering 
$SN \to SN$ and ${\cal A}_N$ is the relevant
matrix element.  For the case of non-relativistic nucleons the singlet-nucleus 
and singlet-nucleon
elastic scattering cross sections are related by \cite{Burgess:2000yq}
\begin{eqnarray}
\sigma^{\rm scalar}_{\rm nucleus} 
&=& \frac{A^2 m^2_r({\rm nucleus},S)}{ m^2_r({\rm nucleon},S)} 
\sigma^{\rm scalar}_{\rm nucleon}
\end{eqnarray}
Numerically evaluating the matrix element appearing in Eq.\ (\ref{eqcross1})
the singlet-nucleon elastic scattering cross section  can be written
as \cite{Burgess:2000yq}
\begin{eqnarray}
\sigma^{\rm scalar}_{\rm nucleon}
&=&
{(\delta_2)}^2 
{\left(\frac{100 ~{\rm GeV}}{M_h}\right)}^4
{\left(\frac{50 ~{\rm GeV}}{M_S}\right)}^2
(5 \times 10^{-42} {\rm cm^2}) .
\label{eqcross2}
\end{eqnarray}
It is evident from the above equations that the scalar cross section for scalar singlet depends on two couplings $\delta_2$ and $\kappa_2$. Hence the dark matter direct detection rate with scalar singlet as the dark matter candidate also depends on the two couplings $\delta_2$ and $\kappa_2$. The constraint on direct detection rates of dark matter from different experiments thus can also put constraint on the parameter space ($\delta_2$, $\kappa_2$) of scalar singlet dark matter. In the next section we will discuss how some of the recent direct detection experiments of dark matter can put limit on the scalar singlet model of dark matter.

%
\begin{figure}[t]
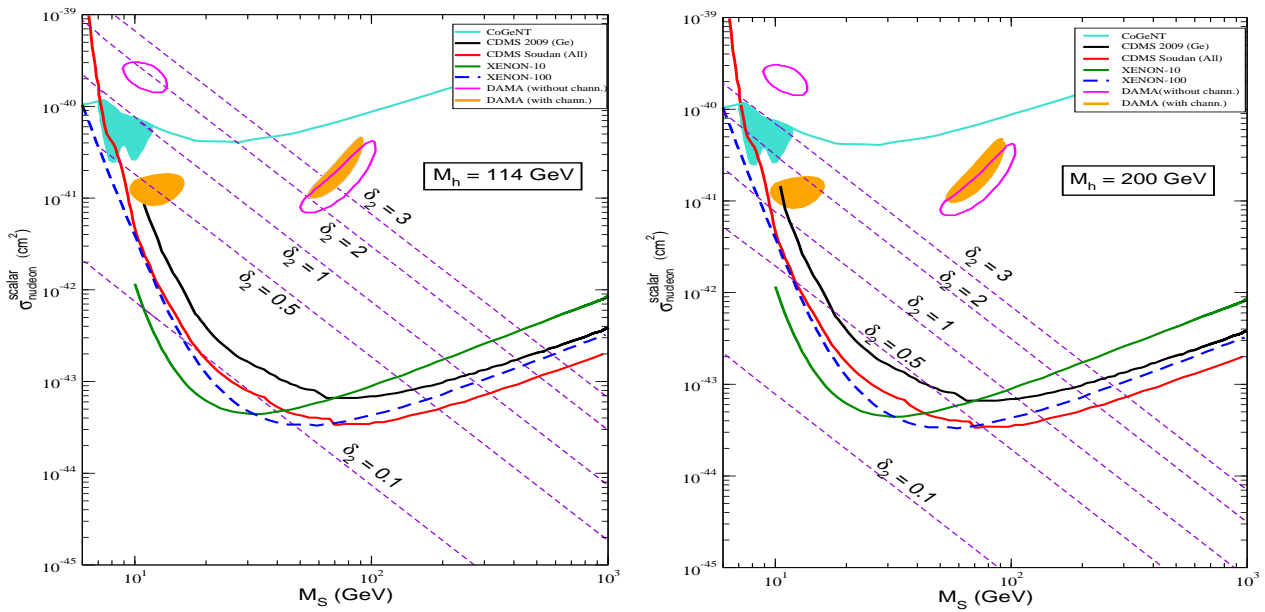

\includegraphics[width=8.2cm, height=8cm, angle=0]{sigvsmd123_114.eps}
\vglue -8.0cm \hglue 8.5cm
\includegraphics[width=8.2cm, height=8cm, angle=0]{sigvsmd123_200.eps}
\caption{\label{fig:svm}
Dashed lines: Scalar singlet-nucleon elastic scattering cross section
($\sigma^{\rm scalar}_{\rm nucleon}$) as a function 
of the singlet mass for different values of $|\delta_2|$ with
2 different values of higgs mass $M_h$, 114 GeV(left panel) and 200 GeV(right 
panel). Solid lines: 90\% C.L. experimental upper limits on 
$\sigma^{\rm DM}_{\rm nucleon}$ from
CDMS 2009 Ge, CDMS Soudan(all), XENON-10 and XENON-100. 
The shaded areas corresponding to 99\% C.L. regions allowed from DAMA 
(with and without channelling) has also been shown.}
\end{figure}
%

\section{Constraining the model parameters}

The direct detection rates of scalar singlet dark matter is governed by 
the scalar-nucleon cross section ($\sigma^{\rm scalar}_{\rm nucleon}$) 
for a given scalar singlet mass $M_S$. Direct detection experiments 
like CDMS, DAMA, Xenon thus can constraint the scalar singlet 
parameter space. Preliminary studies of 
scalar singlet dark matter concerning CDMS, Xenon are given in Refs. 
\cite{Davoudiasl:2004be, Yaguna:2008HD, He:2009YD}. Also, explanation of 
DAMA bounds with scalar singlet dark matter is described 
in \cite{Petriello:2008JJ, Andreas:2008XY}. 
In this work we make a detailed study of all the direct detection 
bounds together with the relic density limits from WMAP and find 
the relevant parameter space for the scalar singlet to be a successful 
dark matter candidate.
 
\subsection{Constraints on $\delta_2 $ as a 
function of dark matter mass}
\label{sec:3.1}

The coupling $\delta_2$ is of severe importance as this is the only coupling in this model which determines the annihilation of the scalar to other standard particles. Moreover given the singlet mass ($M_S$),
$\delta_2$  is the sole coupling that   
controls the singlet-nucleon cross section and the quadratic 
dependence (Eq.\ \ref{eqcross2} ) in particular reflects that 
this cross section is insensitive to the sign of $\delta_2$.

 In Fig.\ \ref{fig:svm} the singlet-nucleon elastic scattering cross section is plotted as a function of singlet mass (dark matter mass) $M_S$ for different values of $\delta_2$. The plots are presented at two different values of Higgs mass ($M_h$), namely 114 GeV (left panel) and 200 GeV (right panel) of Fig. \ \ref{fig:svm}.
For comparison, the 90\% C.L. (confidence limit) results obtained from CDMS II experiment 
(CDMS 2009 (Ge)) \cite{cdms2}
are plotted in Fig.\ \ref{fig:svm} with the similar results from the combined analysis of full data set of Soudan CDMS II results  (CDMS Soudan (All)) \cite{cdms2}, XENON-10 \cite{XENON10} and XENON-100 experiment \cite{XENON100}. 
One sees from Fig.\ \ref{fig:svm} that the DAMA and CoGeNT results are constrained 
in closed allowed regions unlike the other experiments that provide
upper bounds of the allowed masses and cross-sections of dark  matter. 
In this context
it may be noted that the consideration of ion channelling in NaI 
crystal in DAMA 
experiment is crucial in the interpretation of its results. The
presence of channelling in NaI, affects the allowed mass and cross section 
regions of the
DM particles inferred from the observation of an annual modulation 
by the DAMA
collaboration. The effect of channelling has been discussed 
extensively in
\cite{gondolochanneling}. For our analysis we consider the allowed  
mass-cross section limits inferred from observed annual modulation 
of DAMA for both the cases $-$ with channelling and without channelling. 
From Fig.\ \ref{fig:svm} it is seen that for larger values of $\delta_2$,
higher singlet mass domain is required in order to represent the 
experimentally allowed region for  
$\sigma^{\rm scalar}_{\rm nucleon} -$ DM mass plane.  Also, 
the region of overlap of
$\sigma^{\rm scalar}_{\rm nucleon} - M_S$ plots for a 
fixed value of $\delta_2$, 
becomes larger for higher Higgs mass.  

\begin{figure}[t]
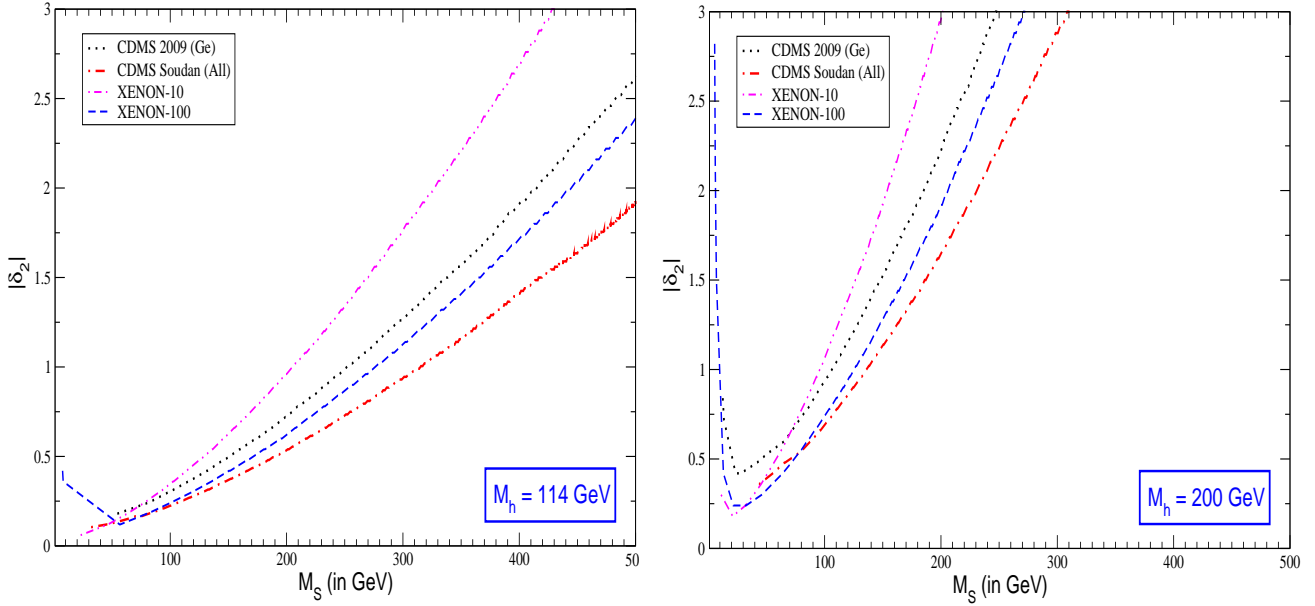

\includegraphics[width=8.5cm, height=8cm, angle=0]{sigdel114.eps}
\vglue -8.0cm \hglue 8.7cm
\includegraphics[width=8.5cm, height=8cm, angle=0]{sigdel200.eps}
\caption{\label{fig:delcon}
Upper limits on $|\delta_2|$ as a function of dark matter mass from 90\%
CL experimental bounds on spin-independent WIMP-nucleon cross section.
Plots are done for 2 values of Higgs mass:
114 GeV (left panel) and 200 GeV (right panel) .}
\end{figure}

In Fig.\ \ref{fig:delcon} we present plots for upper limits of the
range of $|\delta_2|$ (as a function of singlet mass) 
that would reproduce  cross section values 
(computed with Eq.\ \ref{eqcross2})
below the 90\% CL limits of different experiments shown in 
Fig.\ \ref{fig:svm}. It is evident from the plots that, as in 
Fig.\ \ref{fig:svm}, the allowed values of the coupling $ \delta_2 $ 
is sensitive towards the Higgs mass. Moreover the minimum allowed value 
of dark matter mass in this model is also dependent on Higgs mass.
The appearance of local minima at low $M_S$ domain of the plots
are due to the behaviour of experimental bounds that show a  
sudden upturn followed by a rise of the curves at low $M_S$ 
($M_S \to 0$) region.  For
Higgs mass of 120 GeV (114 GeV) this minima corresponds to the  $|\delta_2|$
value $\sim 0.5(0.1)$.
\begin{figure}[t]
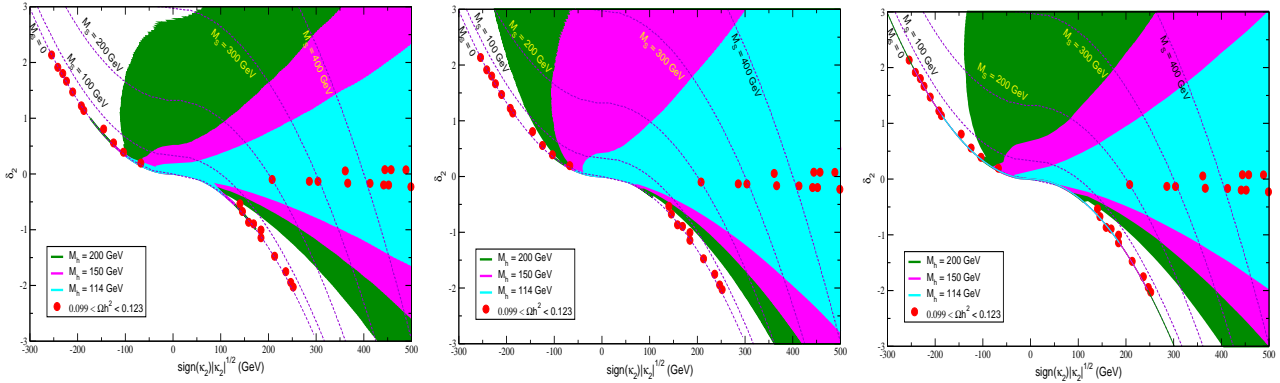

\includegraphics[width=5.5cm, height=5cm, angle=0]{soudanall.eps}
\vglue -5.0cm \hglue 5.7cm
\includegraphics[width=5.5cm, height=5cm, angle=0]{xenon.eps}
\vglue -5.0cm \hglue 11.4cm
\includegraphics[width=5.5cm, height=5cm, angle=0]{xenon100.eps}
\caption{\label{fig:dk1} Shaded region: Range of the parameter space 
$\delta_2 - {\rm sign}(\kappa_2)|\kappa_2|^{1/2}$ 
consistent with 90\% CL limits of WIMP-nucleon scattering cross section
from CDMS Soudan (All)  (left panel), XENON-10 (middle panel) 
and XENON-100 (right panel)
corresponding to three different values of Higgs mass - 
114 GeV, 150GeV, and 200 GeV.
Dashed lines: Different iso-$M_S$ contours in 
$\delta_2 - {\rm sign}(\kappa_2)|\kappa_2|^{1/2}$ plane. The region
spanned by red dots describe the values of model parameters
consistent with WMAP measurements of dark matter relic density : 
$0.099<\Omega h^2<0.123$. }
\end{figure}
\begin{figure}[t]
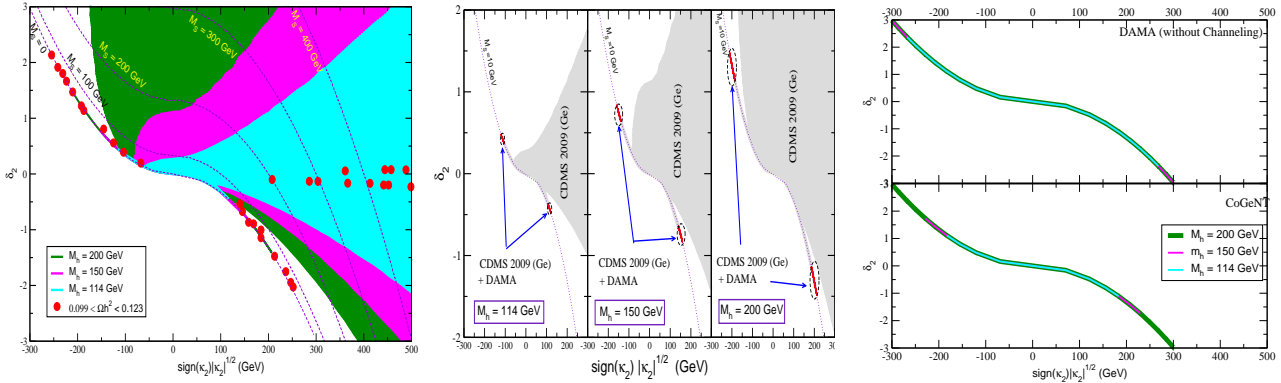

\includegraphics[width=5.5cm, height=5cm, angle=0]{cdms2009.eps}
\vglue -5.0cm \hglue 5.7cm
\includegraphics[width=5.5cm, height=5cm, angle=0]{cdmsdama.eps}
\vglue -5.0cm \hglue 11.4cm
\includegraphics[width=5.5cm, height=5cm, angle=0]{damacogent.eps}
\caption{\label{fig:cdmsdama}
Left panel: (Shaded region) Range of the parameter space 
$\delta_2 - {\rm sign}(\kappa_2)|\kappa_2|^{1/2}$ 
consistent with 90\% CL limits of WIMP-nucleon scattering cross section
from CDMS 2009 (Ge) for three different values of Higgs mass 114 GeV, 150 GeV
and 200 GeV. Different iso-$M_S$ contours in 
$\delta_2 - {\rm sign}(\kappa_2)|\kappa_2|^{1/2}$ plane are shown
by dashed lines.The region
spanned by red dots describe the values of model parameters
consistent with WMAP measurements 
of dark matter relic density : 
$0.099<\Omega h^2<0.123$. Middle panel: (small thread-like regions outlined by
dashed lines) Region in parameter space
consistent with (CDMS 2009 Ge + DAMA(with channelling)) limits for $M_h$ = 114 GeV ($1^{st}$ column), 150 GeV ($2^{nd}$ column), 200 GeV ($3^{rd}$ column). The region 
allowed from only CDMS 2009 (Ge) limits are shaded for corresponding values of
Higgs mass. The contour for $M_S = 10$ GeV are shown by dotted lines. Right panel: Range of the parameter space 
$\delta_2 - {\rm sign}(\kappa_2)|\kappa_2|^{1/2}$ 
consistent with 90\% CL limits of WIMP-nucleon scattering corresponding to the lower DM mass regime allowed from DAMA (without channelling) (upper panel) 
and CoGeNT (lower panel) for three different values of Higgs mass 114 GeV, 
150 GeV and 200 GeV}
\end{figure}

\subsection{Constraining the $\delta_2-\kappa_2$ parameter space}

In the previous subsection, constraining of the parameter $\delta_2$
using other direct detection experimental results are addressed. 
But as described earlier, the present dark matter model with scalar 
singlet also depends on the other parameter namely $\kappa_2$ 
(Eq.\ (\ref{eqms})).  
From  Eq.\ (\ref{eqms}) one sees that the singlet mass term $M_S$  
is governed by two model
parameters namely $\delta_2$ and $\kappa_2$.
For a given value of $\delta_2$, real values of $M_S$ can be obtained 
only by
excluding all $\kappa_2$ values less than $-\delta_2v^2/2$. 
In principle any value of dark matter mass $M_S$ can accommodate
all $(\delta_2,\kappa_2)$ values satisfying 
$\kappa_2 + \delta_2v^2/2 = {M_S}^2$.
However at large $\kappa_2$ values  with
$\kappa_2 \gg \delta_2 v^2$, the singlet mass is predominantly $\kappa_2$
driven and is scaled with it as $\sqrt{\kappa_2}$. The interplay between 
$\delta_2$ and $\kappa_2$ in setting a given singlet mass is represented
in Fig.\ \ref{fig:dk1} where we plotted  (dashed lines) different
iso-$M_S$ contours in 
$\delta_2 - {\rm sign}(\kappa_2)|\kappa_2|^{1/2}$ plane.
For a given singlet mass $M_S$, 
the  range of $|\delta_2|$ consistent with CDMS II/XENON-10 limits
on WIMP-nucleon scattering cross section (as
discussed in Sec.\ \ref{sec:3.1}) would then correspond to a segment
of the corresponding iso-$M_S$ contour in  
$\delta_2 - {\rm sign}(\kappa_2)|\kappa_2|^{1/2}$ plane.
Its projection on the $\kappa_2$ axis 
would give the corresponding range
of the parameter $\kappa_2$. In Fig.\ \ref{fig:dk1} 
the shaded regions represent the domains of the model-parameter space 
$\delta_2 - {\rm sign}(\kappa_2)|\kappa_2|^{1/2}$, that is consistent with 90\% CL limits on WIMP-nucleon elastic 
scattering cross sections from analysis of 
CDMS Soudan (All) (left panel), XENON-10 (middle panel) and XENON-100 (right panel)
results. The dependence of the allowed model-parameter space
on Higgs mass is shown by plotting the allowed areas 
for three different values
of Higgs mass namely 114 GeV, 150 GeV and 200 GeV.  
The region of the parameter
space to the left of
$M_S = 0$ line is excluded as points ($\delta_2,\kappa_2$)
in that region would give negative values of mass. Moving from lower $M_S$ to
higher $M_S$ domain in the parameter space
allows more and more room for $(\delta_2,\kappa_2)$
to represent DM-nucleon scattering cross section 
consistent with its experimental bounds - a feature also apparent
from Fig.\ \ref{fig:svm} and Fig.\ \ref{fig:delcon} (discussed
in Sec.\ \ref{sec:3.1}). At very low scalar singlet mass regime
($M_S \sim 0-10$ GeV) higher peaked value of the WIMP-nucleon
scattering cross section-limit concedes
a thread like extension of the allowed parameter space along 
the corresponding iso-$M_S$ contours. The sudden drop of the 
experimental limits of WIMP-nucleon cross sections 
in $10-20$ GeV mass
regime severely restricts the width of the above mentioned 
thread-like extension of
the allowed parameter space at $M_S \sim 0 - 10$ GeV.
This drop is more robust for the case of the experiment denoted
in this work as ``CDMS Soudan (All)" 
than that of ``XENON" experiment leading to
a more prominent appearance of the thread-like zone
in left panel of Fig.\ \ref{fig:dk1}. The $\delta_2 - \kappa_2$ 
parameter zone obtained from WMAP constraint \cite{Barger:2007im} 
($0.099<\Omega h^2 <0.123$, $\Omega$ being the dark matter relic density 
and $h$ is the Hubble parameter normalized to 100 Km sec$^{-1}$ Mpc$^{-1}$)
are also shown by the red colored dots in Fig. 4, for comparison.
From  Fig.\ \ref{fig:dk1} it is seen that
CDMS/XENON upper limits of WIMP-nucleon scattering cross section
together with WMAP observation allows only small $|\delta_2|$ regime
($\la 0.2$) of the model parameter space for different higher values
of $M_S$ , although for very small $M_S$ values ($0-10$ GeV) CDMS limit
concedes more room for $|\delta_2|$ (upto $\sim 1.0$)

The results of DAMA experiment restricts the variations of  
$\sigma^{\rm scalar}_{\rm nucleon}$ with DM mass in two small 
contours represented as shaded areas in Fig.\ \ref{fig:svm}.
They represent the 99\% C.L.
regions in ($\sigma^{\rm scalar}_{\rm nucleon}$- DM mass) space.
Interpretation of DAMA results with channelling (without channelling) requires
WIMP-nucleon scattering cross sections of order 
$10^{-41} {\rm cm}^2(10^{-40} {\rm cm}^2)$  along with two locally preferred zones of 
dark matter mass -
one around  $\sim 12$ GeV (referred as DAMA-a in this work) and the other 
around $\sim 70$ GeV (referred as DAMA-b in this work).
The DAMA solution corresponding to
the large DM mass regime (DAMA-b) is completely
excluded by observed limits on the WIMP-nucleon scattering
cross section form other direct detection 
experiments like CDMS, XENON etc. The other DAMA solutions corresponding to
lower DM mass regime (DAMA-a) are also largely disfavoured by
XENON. However though a small region at the lower DM mass 
regime (``with channelling" case)
in DAMA-a zone is barely 
consistent with 90\% C.L. with CDMS experiment denoted in this work as 
``CDMS 2009 (Ge)", the corresponding limit ``CDMS Soudan (All)" 
experiment excludes the entire DAMA-a zone.
In the left panel of Fig.\ \ref{fig:cdmsdama} we have shown region of 
$\delta_2 - {\rm sign}(\kappa_2)|\kappa_2|^{1/2} $
parameter space that corresponds to singlet-nucleon elastic
scattering cross sections within its 90\% C.L. limit from 
CDMS 2009 (Ge). The middle panel of Fig.\ \ref{fig:cdmsdama} shows
the $\delta_2 - {\rm sign}(\kappa_2)|\kappa_2|^{1/2} $ parameter space, consistent with both CDMS 2009 (Ge) and DAMA results (with channelling).  
The small thread like regions (marked red and 
indicated within closed dashed lines) 
in the middle panel of Fig.\ \ref{fig:cdmsdama}
represent the parameter space domain that fits singlet-nucleon
scattering cross section with DM-nucleon cross section consistent with 
both CDMS 2009 (Ge) and DAMA limits (with channelling). These are presented 
in three
different columns that correspond to three
values of Higgs mass namely 114 GeV, 150 GeV and 200 GeV. 
The parameter space consistent with CDMS 2009 (Ge) limit
only for corresponding values of Higgs masses
are also shown in the respective columns by grey shades.
The iso-$M_S$ contour for $M_S = 10$ GeV, spanning through 
the DAMA + CDMS 2009 (Ge) allowed regimes, is 
also shown in each column of the same figure to illustrate the fact that 
parameter space regions 
consistent with both DAMA and CDMS 2009 (Ge) correspond
to a singlet dark matter mass of around $\sim 10$ GeV. In the right panel of 
Fig.\ \ref{fig:cdmsdama} we have shown the allowed region in 
the $\delta_2 - {\rm sign}(\kappa_2)|\kappa_2|^{1/2} $ parameter space
consistent with DAMA-a region without channelling (upper panel) and the
CoGeNT result (the lower panel).

\section{Predictions for Rates at Argon detector}
The Argon detector is a Noble liquid detector where the liquefied noble gas
Argon is used as targets for direct detection of WIMPs. Because of the high density
and high atomic number the event rate is expected to be large. Also
because of its high scintillation and ionization yields owing to its low
ionization potentials, it can effectively discriminate nuclear recoils 
and other backgrounds from $\gamma$ or electrons. The ArDM (Argon
Dark Matter) experiment at surface of CERN uses one such detector
that envisages one ton of liquid Argon in a cylindrical container. This has 
a provision for three dimensional imaging for every event. 
A strong electric field along the axis of the cylinder helps drifting
of the charge -- produced due to the ionization of liquid Argon by WIMP 
induced nuclear recoil -- to the surface of the liquid.
This charge then enters into the gaseous phase of
the detector (TPC) where it is multiplied through avalanche and finally recorded
by a position sensitive readout. \\

In this section we estimate WIMP signal rates for such Argon detector. 
The differential rate for dark matter scattering
detected per unit detector mass can be
written as
\begin{eqnarray}
\frac {dR} {d|{\bf q}|^2} = N_T \Phi \frac {d\sigma} {d|{\bf q}|^2} \int f(v) dv
\label{eq:drdqsq}
\end{eqnarray}
where $N_T$ is number of target nuclei per unit mass of the detector, $\Phi$
is the dark matter flux, $f(v)$ denotes the distribution of
dark matter velocity $v$ (in earth's frame). The integration
is over all possible kinematic configurations in the scattering process.
$|\bf q|$ is the momentum transferred to the nucleus in
dark matter-nucleus scattering and $\sigma$ being the corresponding
cross section. The recoil energy of the scattered nucleus can be expressed in 
terms of the momentum transfer $|\bf q|$ as
\begin{eqnarray}
E_R 
&=& |{\bf q}|^2/2M_N 
= m^2_r v^2 (1 - \cos\theta)/M_N
\label{eq:recoil}
\end{eqnarray}
where $M_N$ is the nuclear mass, $\theta$ is the scattering angle
in dark matter - nucleus center of momentum frame and $m_r$ is the reduced
mass given by
\begin{eqnarray}
m_r &=& \frac{M_N M_{S}}{M_N + M_{S}}\,\,.
\end{eqnarray}
In the above $M_{S}$ is the dark matter mass. Expressing 
the dark matter
flux $\Phi$ in terms of the local dark matter density $\rho_s$,
velocity $v$ and mass $M_{S}$. With $N_T = 1/M_N$ 
and writing $|{\bf q}|^2$ in terms
of nuclear recoil energy $E_R$, Eq.\ (\ref{eq:drdqsq}) can be rewritten as
\begin{eqnarray}
\frac {dR} {dE_R}
&=&
2 \frac {\rho_{s}} {M_{S}} \frac {d\sigma}
{d |{\bf q}|^2} \int_{v_{min}}^\infty v f(v) dv
\label{eq:drde}
\end{eqnarray}
where
\begin{eqnarray}
v_{\rm min} &=& \left [ \frac {M_{N} E_R} {2m^2_{\rm r}} \right ]^{1/2}
\end{eqnarray}
The dark matter - nucleus differential cross-section for the scalar interaction 
is given by \cite{Jungman:1995df}
\begin{eqnarray}
\frac {d\sigma} {d |{\bf q}|^2} = \frac {\sigma^{\rm scalar}}
{4 m_{\rm red}^2 v^2} F^2 (E_R) \,\,\, .
\label{eq:dsdqsq}
\end{eqnarray}
Here $\sigma^{\rm scalar}$ is dark matter-nucleus scalar cross-section
and $F(E_R)$ is nuclear form factor given by \cite{helm,engel}
\begin{eqnarray}
F(E_R) &=& 
\left [ \frac {3 j_1(q R_1)} {q R_1} \right ] 
{\rm exp} \left ( \frac {q^2s^2}
{2} \right ) \\
R_1 &=& (r^2 - 5s^2)^{1/2} \nonumber \\
r &=& 1.2 A^{1/3} \nonumber
\end{eqnarray}
where $s (\simeq 1 ~{\rm fm})$ is the thickness parameter
of the nuclear surface, $A$ is the mass number of the nucleus,
$j_1(qR_1)$ is the spherical
Bessel function of index 1 and $q = {\bf |q|} = \sqrt{2M_NE_R}$ as from
Eq. \ (\ref{eq:recoil}). Assuming distribution $f(v_{\rm gal})$
of dark matter velocities $(v_{\rm gal})$ 
with respect to galactic rest frame to be Maxwellian, 
one can obtain the distribution  $f(v)$ of dark matter velocity ($v$) with respect to earth rest frame by
making the transformation
\begin{eqnarray}
{\bf v} = {\bf v}_{\rm gal} - {\bf v}_\oplus
\end{eqnarray} 
where ${\bf v}_\oplus$ is the velocity 
of earth with respect to Galactic rest
frame and is given as a function of time $t$ by
\begin{eqnarray}
v_\oplus &=& v_\odot + v_{\rm orb} \cos\gamma 
\cos \left (\frac {2\pi (t - t_0)}{T} \right ) \label{eq:vearth}
\end{eqnarray}
In Eq.\ (\ref{eq:vearth}) $v_\odot$ is the speed of the solar system
in galactic rest frame, $T$ ($1$ year) us the period of earth's rotation
about sun , $t_0 = 2^{\rm nd}$ June (the time of the year when the
orbital velocity of earth and velocity of solar system 
point in the same direction) and $\gamma \simeq 60^o$ is the angle 
subtended by earth orbital
plane at Galactic plane. The speed of solar system $v_\odot$ in the
Galactic rest frame is given by
\begin{eqnarray}
v_\odot &=& v_0 + v_{\rm pec}
\end{eqnarray}
$v_0$ being the circular velocity of the local system at the position of
solar system and $v_{\rm pec} = 12 {\rm km/sec}$, called  
{\it peculiar velocity},
is speed of solar system with respect to
the local system. Physical range of $v_0$ is $170\,\, {\rm km/sec} \leq v_0 \leq 270$ km/sec (90 \% C.L.) \cite{pec1,pec2}. In this work we consider the central value -  $220$ km/sec  for   $v_0$. The term
$\cos2[\pi (t - t_0)/T]$ in the velocity is responsible for annual modulation
of dark matter signal. Introducing a dimensionless quantity $T(E_R)$ as
\begin{eqnarray}
T(E_R) 
&=& \frac {\sqrt {\pi}} {2} v_0 \int_{v_{\rm min}}^\infty \frac {f(v)}
{v} dv\,\,
\end{eqnarray}
which can also be expressed as 
\begin{eqnarray}
T(E_R) = \frac {\sqrt {\pi}} {4v_\oplus} v_0 
\left [ {\rm erf} \left ( \frac{v_{\rm min} + v_\oplus} {v_0} \right ) 
-  {\rm erf} \left ( \frac
{v_{\rm min} - v_\oplus} {v_0} \right ) \right ]
\end{eqnarray}
we obtain from Eqs.\ (\ref{eq:drde}) and (\ref{eq:dsdqsq})
\begin{eqnarray}
\frac {dR} {dE_R} 
&=& \frac {\sigma^{\rm scalar}\rho_s} {4v_\oplus M_S
m_{\rm r}^2} F^2 (E_R) 
\left[
{\rm erf}\left(\frac{v_{min} + v_\oplus}{v_0}\right)
- {\rm erf}\left(\frac{v_{min} - v_\oplus}{v_0}\right)
\right]
\end{eqnarray}
%
\begin{figure}[t]
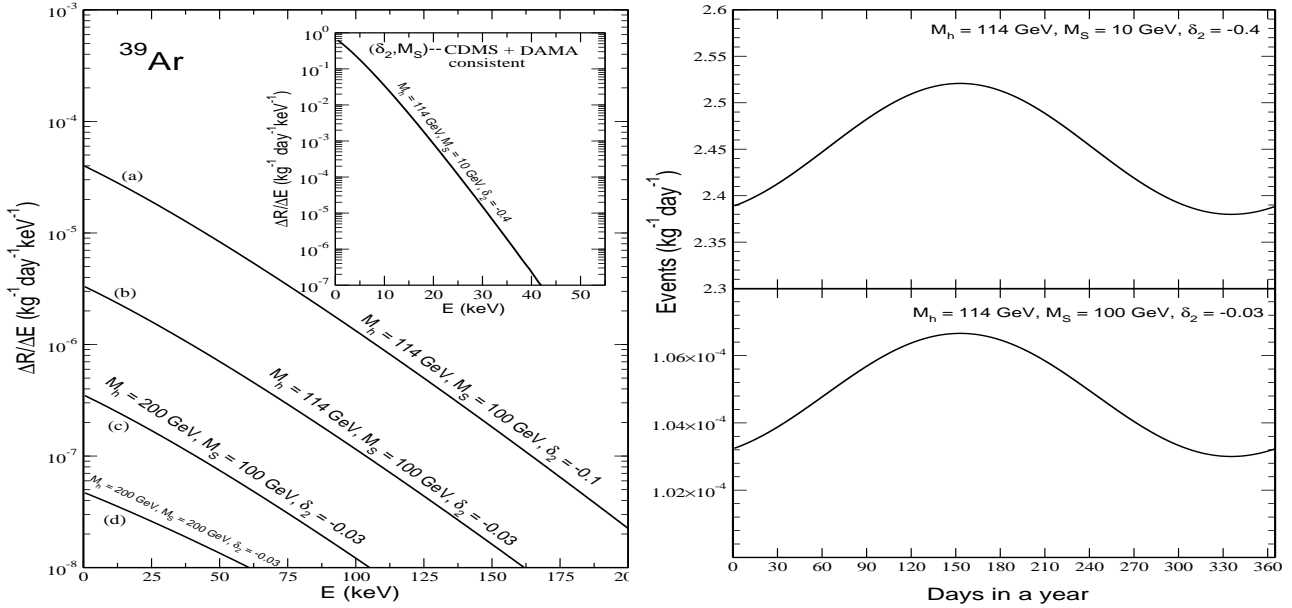

\includegraphics[width=8.5cm, height=8cm, angle=0]{ratear.eps}
\vglue -8.0cm \hglue 8.5cm
\includegraphics[width=8.5cm, height=8cm, angle=0]{annvarar.eps}
\caption{\label{fig:rate}
Left Panel: Plot of predictions for dark matter detection rates (per kg per day per keV)
in Argon detector as a function of observed recoil energy. The plots are shown
for two values of $M_h$ - $114$ GeV and $200$ GeV and for two sets of 
($M_S, \delta_2$) values which are consistent with CDMS limits as well as
observed relic density of dark matter (WMAP). In the inset we show 
the corresponding
plot for $M_h = 114$ GeV and for two sets of ($M_S, \delta_2$) values simultaneously
consistent with limits on scattering cross section from CDMS 2009 (Ge) and DAMA 
and also with WMAP. Right panel: Predicted annual
variation of event rates in Argon detector over one year for $M_h = 114$ (GeV).
The upper panel corresponds to a ($M_S, \delta_2$) value consistent
with CDMS+DAMA+WMAP while the lower panel corresponds to a 
($M_S, \delta_2$) value consistent with CDMS+WMAP.}
\end{figure}
The local dark matter
density $\rho_s$ may be taken as 0.3 GeV/cm$^3$.
The observed recoil energy $(E)$ in the
measured response of the detector is a 
fraction $(Q_X)$ of actual recoil energy $(E_R)$
at the time of scattering. This fraction $Q_X = E/E_R$ (called as
quenching factor) is different for different scattered nucleus $X$.
For  $^{39}$Ar, $Q_{\rm Ar}$ = 0.76.
Thus the differential rate in terms of the observed recoil energy 
$E$ for $^{39}Ar$ detector 
can be expressed as 
\begin{equation}
\frac {\Delta R} {\Delta E} (E) =
\displaystyle\int^{(E + \Delta E)/Q_{\rm Ar}}_{E/Q_{\rm Ar}}
\frac {dR_{\rm Ar}} {dE_R} (E_R) \frac {\Delta E_R} {\Delta E}
\end{equation}
In left panel of Fig.\ \ref{fig:rate} we show the expected differential
rates (/kg/day/keV) for different observed recoil energies in Argon detector
considering scalar singlet as the dark matter candidate.
Four representative cases have been plotted and they are
denoted as (a), (b), (c) and (d). For all the plots, (a) - (d), the chosen 
values of coupling $\delta_2$ (as also corresponding scalar singlet masses, $M_S$)
are consistent with current CDMS and WMAP limits. 
All the plots show that the 
rate falls off with the increase of recoil energy. 
Plots (a) and (b) are the variations of rates for the same set of 
Higgs mass ($M_h = 114$ GeV) and singlet mass ($M_S = 100$ GeV)
but for different values of the coupling $\delta_2$ ($\delta_2 = -0.1 (-0.03)$ 
for plot (a) ((b)). Plots (a) and (b) show a decrease of the rate when 
$|\delta_2|$ decreases. For example, in case of recoil energy $E = 50$
GeV the calculated rates from plots (a) and (b) 
are $8.5 \times 10^{-6}$ (for   $|\delta_2| = 0.1$)
and $7.2 \times 10^{-7}$ (for $|\delta_2| = 0.03$ ) respectively in the units 
of /kg/day/keV. 
Plots (c) and (d) compare the variation of rates for two different
scalar masses namely $M_S = 200$ GeV (plot (c)) and $M_S = 100$ GeV 
(plot (d)) for same values of $M_h, \ \delta_2$ (200 GeV, $-0.03$). From 
plots (c) and (d) it is seen that the rate increases for any particular value of 
recoil energy with decrease of dark matter mass. 
For example in case of $E = 50$ GeV, the rates for $M_S= 100$ GeV (plot c) 
and $M_S  = 200$ GeV (plot d) the calculated rates are $8 \times 10^{-8}$
and $1.4 \times 10^{-8}$ respectively in the units of /kg/day/keV.
This is evident from the 
expression for scalar cross section (Eq.\ (\ref {eqcross2}) which  
varies as $M_S^{-2}$ and direct detection rates is linear with the scalar cross section.
One can compare plots (b) and (d) to see the effect of Higgs mass values
on the rate. For  $M_S, \ \delta_2$ (100 GeV, $-0.03$) , the estimated rates in the 
present calculations  at 
$E = 50$ keV are $7.2 \times 10^{-7}$ (/kg/day/keV) for $M_h = 114$ GeV (plot (b))
which is reduced to $1.4 \times 10^{-8}$ (/kg/day/keV) for $M_h = 200$
GeV (plot (d)).  In the inset of this 
same figure we show the calculated prediction of rates for 
$M_S = 10$ GeV, $\delta_2 =  0.4$. This is compatible with 
the CDMS and DAMA bound together, other than satisfying the WMAP 
limits.  The Higgs mass is kept at 114 GeV. As an example, 
for $E = 30$ GeV the calculated expected event is 
$1.4 \times 10^{-3}$ per day for a ton of the detector.

     
One very positive signature of dark matter in direct detection method
is the annual variation of detection rate. This periodic variation arises due
to periodic motion of the earth about the sun in which the directionality 
of earth's motion changes continually over the year. As a result, there is
an annual variation in the amount
of dark matter encountered by the earth. The detection of annual variation
in direct detection experiments serve as a smoking gun signal for existence
of dark matter.
In right  panel of Fig.\ \ref{fig:rate} we show the calculated annual variation 
of event rate (/kg/day) in different times of a year.  In the upper panel,
we have chosen the parameter set that is compatible with 
CDMS, DAMA and WMAP limits (as in inset of the left panel) and 
for the lower panel we have given the results for 
$M_h = 114$ GeV, $M_S = 100$ GeV and $\delta_2 = -0.03$.
As expected,
the yield is maximum on 2nd June when the direction of motion of the earth 
is the same as that of the solar system.

\section{Summary and Conclusions}
In the present work we  consider a simplest extension of SM, introducing 
a real scalar singlet along with a discrete $Z_2$ symmetry
which ensures  stability of the singlet. Such  singlet is considered as a
viable cold dark matter candidate.
The scattering of this singlet dark matter off nuclei  in the detector 
can be observed by measuring the  energy of the recoil nuclei.
The calculated singlet-nucleon scattering cross section in this model
explicitly depends on the coupling $\delta_2$ and implicitly on $\kappa_2$
(as defined in Eq.\ (\ref{eqpot})).  
We constrain the $\delta_2 - \kappa_2$ 
parameter space using the recent bounds on the 
WIMP-nucleon scalar cross section as function of WIMP mass, 
reported by the CDMS  collaboration and also reported by the XENON-10
and XENON-100 collaborations. 
The allowed zones in  the parameter space follow a typical pattern determined
by the shape of the WIMP mass dependence of the experimental limits 
considered here 
and the way the scalar singlet mass 
is related to  $\delta_2$ and $\kappa_2$. The allowed zones vary for different
Higgs masses and they are consistent  with WMAP limits.

We investigate the effect of the inclusion of DAMA results (with 
channelling) on the 
 $\delta_2 - \kappa_2$ parameter space but the allowed zone is found to 
be extremely small and representative of a very low dark 
matter mass (around $\sim 10$ GeV). We also compute the range of 
$\delta_2 - \kappa_2$ parameter space consistent with 
DAMA (without channelling) and CoGeNT experiments.

Utilising the constrained parameter space we estimate the possible 
detection rates and their annual variations for a liquid Argon detector.\\

{\bf Acknowledgments:} We thank Probir Roy and Biswajit Adhikary
for useful discussion. S.C. acknowledges support from the projects 
‘Centre for Astroparticle Physics’ and ‘Frontiers of Theoretical Physics’
of Saha Institute of Nuclear Physics. A.G. and D.M. acknowledge 
the support from the DAE project 
`Investigating Neutrinoless Double Beta Decay, Dark Matter and GRB'
 of Saha Institute of Nuclear Physics.


\end{document}